\documentclass[twocolumn,english,prl,showpacs]{revtex4}
\usepackage[T1]{fontenc}
\usepackage[latin9]{inputenc}
\usepackage{amssymb}
\usepackage{graphicx}
\usepackage{amsmath,color}
\usepackage{mathrsfs}
\usepackage{float}
\usepackage{indentfirst}

\makeatletter

\def\journal #1, #2, #3, 1#4#5#6{{\sl #1~}{\bf #2}, #3 (1#4#5#6) }

\newcommand{\tmop}[1]{\ensuremath{\operatorname{#1}}}

\makeatother

\begin{document}

\title{Interaction induced topological phase transition in  Bernevig-Hughes-Zhang model}

\author{Lei Wang$^{1}$, Xi Dai$^2$ and X. C. Xie$^{3}$}

\affiliation{$^{1}$ Theoretische Physik, ETH Zurich, 8093 Zurich, Switzerland}

\affiliation{$^{2}$Beijing National Lab for Condensed Matter Physics and Institute of Physics, Chinese Academy of Sciences, Beijing 100190, China }

\affiliation{$^{3}$International Center for Quantum Materials and School of Physics, Peking University, Beijing 100871, China}


\begin{abstract}
We study interaction induced topological phase transition in Bernevig-Hughes-Zhang model. Topological nature of the phase transition is revealed by directly calculating the Z$_{2}$ index of the interacting system from the single-particle Green's function. 
The interacting Z$_{2}$ index is also consistently checked through the edge spectra.
Combined with \textit{ab initio} methods, present approach is a useful tool searching for correlated topological insulating materials from the first-principle point of view.
\end{abstract}






\maketitle

\begin{figure}[t]
\begin{center}
\includegraphics[width=0.4\textwidth]{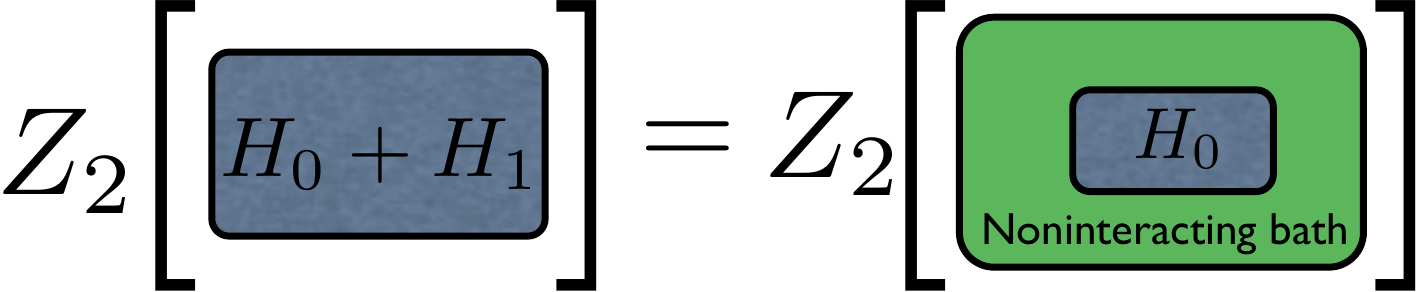}
\caption{Illustration of strategy for calculating the interacting Z$_{2}$ index.  Z$_{2}$ index of an \emph{interacting} system equals to Z$_{2}$ index of a \emph{noninteracting} auxiliary system \cite{Wang:2011p39810}.
The auxiliary system is constructed such that its noninteracting bath mimics the effect of local self-energy due to the interaction $H_{1}$. See text for details on how to construct the auxiliary system.
}
\label{fig:PE}
\end{center}
\end{figure}

\textit{Introduction--}Ab initio methods got great successes in predicting topological insulating materials thus far\cite{Zhang:2009p7135,Zhang:2011p30949,Xiao:2010p36216, Zhang:2011p28394}. One reason behind is that correlation effect is relatively weak in relevant materials. Another reason is that the topological property is more robust than the magnetic or the superconducting properties against systematical errors in the first-principle calculations. Study the interplay of the correlation effect and the topological order is interesting from both fundamental and application points of view. Interacting effect on topological phases already received many attentions recently. Various analytical or numerical tools have been applied to several models for topological insulators\cite{Varney:2010p21916, Yu:2011p35101, Hohenadler:2011p28789,Yoshida:2011p44360, Wang:2010p25724}. However, most of these studies used the conventional non-topological quantities such as the long range order parameter\cite{Rachel:2010p20458}, the excitation gap\cite{Hohenadler:2011p28789, Varney:2011p45724} {\it etc} to indirectly characterize the interacting topological phases. How to extract the interacting topological index from many-body calculations remains to be a challenge.

There are indeed some progresses along this line. For Chern TI, the quantum Hall conductance can be calculated using the twisted-boundary-condition \cite{Niu:1985p19667} or the linear response formula\cite{Yoshida:2011p44360}. For Z$_{2}$ TI, Ref.\cite{Wang:2010p25171} suggested a definition of interacting Z$_{2}$ index based on the single-particle Green's function:

\begin{eqnarray}
 Z_{2} & =  & \frac{\pi^2}{15} \varepsilon_{\mu \nu \rho\sigma\tau } \mathrm{Tr} \int  \frac{\mathrm{d}^{4}k \mathrm{d}\omega}{(2\pi)^{5}} \, \nonumber \\
  & &G \partial_{\mu} G^{- 1} G \partial_{\nu} G^{- 1} G \partial_{\rho} G^{- 1}G \partial_{\sigma} G^{- 1}G \partial_{\tau} G^{- 1}
\label{eqn:ishikawa}
\end{eqnarray}
where $G({\mathbf k},i\omega)$ is Matsubara Green's function \footnote{The system should be gapped, otherwise integrating out fermions is invalid and in general there is no quantized topological index.}, $\varepsilon_{\mu\nu\rho\sigma\tau}$ is the five-order anti-symmetric tensor, $\mu, \nu...\tau$ indices denote the frequency-momenta $(\omega, k_x, k_y, k_{z}, k_{\lambda})$. For three (two) dimensional topological insulator, one (two) of them is  pumping parameter extending the momenta to four dimensions \cite{Wang:2010p25171}. Note the difference of Eqn.\ref{eqn:ishikawa} with quantum (spin) Hall conductivity coefficient \cite{Yoshida:2011p44360}. Since the formula contains the five-dimensional frequency-momenta integration and here  the conservation of spin quantum number is not assumed, hence Green's function is in general a matrix in spin-orbital space. Although being very general, introducing of pumping parameter and invoking of derivatives with respect to frequency-momenta hinder precise numerical implementation of Eqn.\ref{eqn:ishikawa}.

There are some attempts of integrating Eqn.\ref{eqn:ishikawa} numerically for typical forms of interacting Green's functions \cite{Wang:2011p43509}. 
It is highly desirable to have a recipe for extracting topological informations for Z$_{2}$ TI. It would be even more useful if the method can be easily combined with the first principle approaches. This will accelerate the search for correlated topological materials.

In this Letter, we propose a numerical scheme of calculating interacting Z$_{2}$ index in Eqn.\ref{eqn:ishikawa}. It is based on the pole-expansion form of the local self-energy \cite{Wang:2011p39810}. With interacting Z$_{2}$ index we study an interaction induced topological phase transition and nontrivial topological phases in interacting Bernevig-Hughes-Zhang (BHZ) model\cite{Bernevig:2006p44981}.




As shown in Ref.\cite{Wang:2011p39810}, when the self-energy is local ($\mathbf k$-independent), the Z$_{2}$ index of an interacting system is identical to that of an auxiliary noninteracting system. The auxiliary system is constructed through the pole-expansion technique. The idea of pole-expansion is illustrated in Fig.\ref{fig:PE}. We are interested in the topological properties of an interacting system with Hamiltonian $H_{0}+H_{1}$, where $H_{0}$ is single-particle part, $H_{1}$ describes the local interactions. Suppose effect of the local interaction imprints some local self-energy to the system. One can construct an auxiliary noninteracting system by coupling noninteracting baths to $H_{0}$. Integrating out the bath sites gives identical self-energy due to the local interaction $H_{1}$. Pole-expansion of the local self-energy is an efficient way of determining the bath levels. Most importantly, Ref.\cite{Wang:2011p39810} proves that the Z$_{2}$ index of the \emph{noninteracting} system is identical to the Z$_{2}$ index of the original \emph{interacting} system. This greatly simplifies the study of the topological properties of interacting insulators. Note that this formalism does not require the system holds inversion symmetry.


%
%
%

\textit{Model and Methods--}We study interacting effect on BHZ model, one of the early non-interacting models showing the Z$_{2}$ topological order\cite{Bernevig:2006p44981}. The Hamiltonian reads $H_{0} = \sum_{\mathbf k} \psi_{\mathbf k}^{\dagger} H_{\mathbf k}  \psi_{\mathbf k}$, with  $\psi^{\dagger}_{\mathbf k} = (c_{s{\mathbf k}\uparrow}^{\dagger},c_{p{\mathbf k}\uparrow}^{\dagger}, c_{s{\mathbf k}\downarrow}^{\dagger}, c_{p{\mathbf k}\downarrow}^{\dagger})$, 

\begin{equation}
 H_{\mathbf k}  = \left(\begin{array}{cc}
  \mathcal{H}_{\mathbf k} & \\
  & \mathcal{H}_{- \mathbf k}^{\ast}
\end{array}\right)
\end{equation}

\begin{equation}
\mathcal{H}_{\mathbf k} = \left(\begin{array}{cc}
  \mathcal{M} ({\mathbf k}) & \lambda(\sin k_x - i \sin k_y)\\
  \lambda(\sin k_x + i \sin k_y) & - \mathcal{M} ({\mathbf k})
\end{array}\right)
\end{equation}
and $\mathcal{M} ({\mathbf k}) = m + \cos k_{x}+\cos k_{y}$. Parameter $m$ sets the energy offset between the $s$ and $p$ orbitals. Spin-orbit coupling $\lambda$ induces hybridization between them. Noninteracting BHZ model is in Z$_{2}$ topological phase for $-2<m<2$ (except $m=0$ the gap closes)\cite{Qi:2008p12545}.

We introduce intra-orbital repulsion $H_{1}$, and study its effect on the topological properties of BHZ model.
\begin{equation}
H_{1} = U\sum_{i} (n_{si\uparrow}n_{si\downarrow} + n_{pi\uparrow}n_{pi\downarrow} )-
\frac{U}{2}\sum_{i\sigma} (n_{si\sigma}+ n_{pi\sigma})
\end{equation}
In the following discussion, we focus on $m = -3$ and $\lambda = 0.3$. Chemical potential term ensures that the whole system has particle-hole symmetry and on average there are two particles per site.

For $U = 0$ the system lies in the normal band insulator region. $s$ and $p$ orbitals are well separated. Occupation of the two orbitals are $(2^{-},0^{+})$, small deviation is due to hybridizations. With increasing $U$, some of elections reside on $s$-orbital will be pushed to $p$ -orbital. Occupation number of each orbital will approach to $(1^{+}, 1^{-})$ at large $U$. In following calculations, spin symmetry is conserved,  where we have $\langle n_{s\uparrow}\rangle =\langle n_{s\downarrow}\rangle\equiv \langle n_{s}\rangle$ and $\langle n_{p\uparrow}\rangle =\langle n_{p\downarrow}\rangle\equiv \langle n_{p}\rangle$ .  

We perform the dynamical-mean-field-theory (DMFT) \cite{Georges:1996p5571} calculations for interacting BHZ model. By solving the two-orbital Anderson impurity model (AIM) self-consistently, DMFT gives a good approximation of the local Green's function and self-energy. Here, the AIM is solved with Lanczos exact diagonalization (ED) impurity solver \footnote{We include three sites to represents bath. At each step we determine energy level and coupling strength of bath sites to the impurity site from center of mass of and weight of hybridization function.}. ED solver captures essential features of the correlation and has the advantage of giving us direct access to pole structure of Green's function. We employing Lanczos method \cite{Balzer:2011p44597} to solve ground state and Green's function. With ground state $|g\rangle$ of AIM, we run Lanczos algorithm starting from $ c_{\beta}^{\dagger} |g\rangle$ ($ c_{\alpha} |g\rangle$)  to get a set of eigenpairs $\{E_{n}, |n\rangle\}$ ($\{E_{m}, |m\rangle\}$) for adding (removing) one particle subspace ($\alpha,\beta$ denote combined spin-orbital index $\alpha=\{s(p),\sigma\}$) . Local Green's function is calculated as:

\begin{eqnarray}
G_{\alpha\beta}(z)= \sum_{n} \frac{\langle g|c_{\alpha}| n\rangle \langle n| c_{\beta}^{\dagger} |g\rangle}{z -E_{n}+E_{g}} +   \sum_{m} \frac{\langle g|c_{\beta}^{\dagger}| m\rangle \langle m| c_{\alpha}|g\rangle}{z + E_{m}-E_{g}}
\label{eqn:GF}
\end{eqnarray}

The resulting Green's function is diagonal, \textit{i.e.} $G_{\alpha\beta} = G_{\alpha}\delta_{\alpha\beta}$. Rewrite Eqn.\ref{eqn:GF} to $G_{\alpha}(z) = \sum_{i=1}^{N_{G}} \frac{W^{\alpha}_{i}}{z-Q_{\alpha}^{i}}$, we have direct access to poles and weights of local Green's function $\{Q^{i}_{\alpha},W^{i}_{\alpha}\}$. Number of poles $N_{G}$ are limited by  basis size as well as number of Lanczos vectors. We have checked that finial Z$_{2}$ index does not change as we include more Lanczos vectors. We then get pole-expansion of local self-energy $\Sigma_{\alpha}(z) = \Sigma_{\alpha}(\infty) +  \sum_{i=1}^{N_{G}-1} \frac{V_{\alpha}^{i}}{z-P_{\alpha}^{i}}$ follow Ref.\cite{Zhao:2011p43243}, with

\begin{eqnarray}
\Sigma_{\alpha}(\infty) =  \sum_{i}^{N_{G}} W^{i}_{\alpha} Q^{i}_{\alpha} \\
\sum_{i=1}^{N_{G}} \frac{W^{i}_{\alpha}}{P^{j}_{\alpha}-Q^{i}_{\alpha}} = 0 \\
V^{j}_{\alpha} = \left[\sum_{i=1}^{N_{G}} \frac{W^{i}_{\alpha}}{(P^{j}_{\alpha}-Q^{i}_{\alpha})^{2}}   \right]^{{-1}}
\end{eqnarray}

\begin{figure}[t]
\begin{center}
\includegraphics[width=0.5\textwidth]{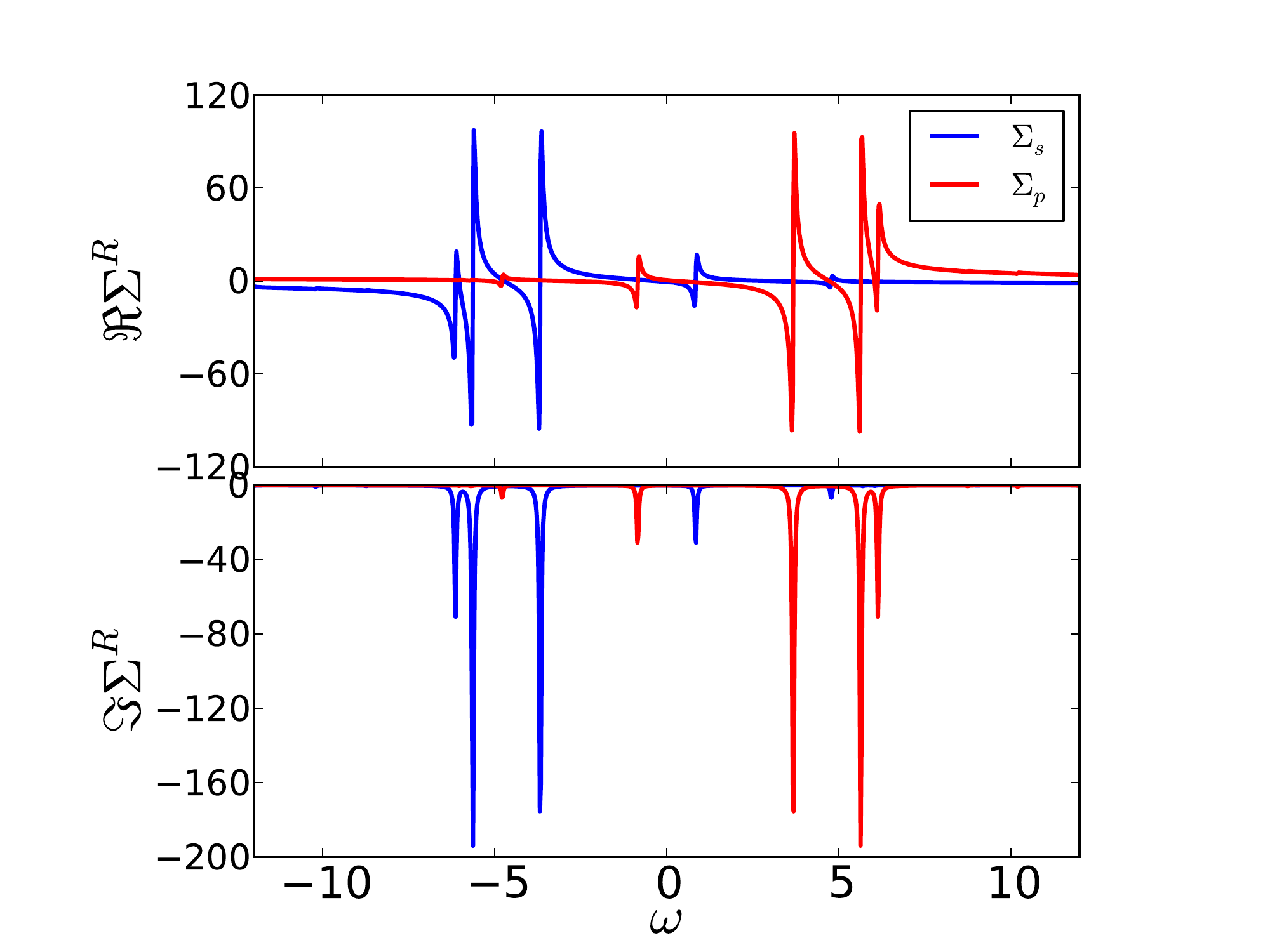}
\caption{Retarded self-energy for $s$ (blue) and $p$ (red) orbitals at $U=8$. Self-energy plot using $\{\Sigma_{\alpha}(\infty),P_{\alpha}^{i},V_{\alpha}^{i}\}$ gives identical curve. Notice that only taking into account of high-frequency tail of self-energy $\Sigma_{\alpha}(\infty)$ will erroneously predict a trivial Z$_{2}$ insulator at $U=8$.
}
\label{fig:Sigma}
\end{center}
\end{figure}

We explicitly check that the self-energies constructed in this way agree with the self-energies obtained directly from the Dyson equation, see Fig.\ref{fig:Sigma} \footnote{Different from Ref.\cite{Zhao:2011p43243} where they employ Hubbard-I solver, here we need to further exclude poles and corresponding weights from hybridization with bath sites. This can be easily done since we have discrete bath sites in ED solver, whose hybridization function is known.}. With $\{\Sigma_{\alpha}(\infty),P_{\alpha}^{i},V_{\alpha}^{i}\}$ we construct the noninteracting pseudo-Hamiltonian according to \cite{Savrasov:2006p3872, Wang:2011p39810}:

\begin{eqnarray}
  \tilde{H}_{\mathbf k}  =  \left(\begin{array}{cccc}
    H_{\mathbf k} + \mathrm{diag} (\Sigma_{\alpha} (\infty)) & \mathrm{diag} (
    \sqrt{V_{\alpha}^1}) & \ldots & \mathrm{diag} ( \sqrt{V_{\alpha}^{N_G -
    1}})\\
    \tmop{diag} ( \sqrt{V_{\alpha}^1}) & \mathrm{diag} (P^{1}_{\alpha}) &  & \\
    \vdots &  & \ddots & \\
    \mathrm{diag} ( \sqrt{V_{\alpha}^{N_G - 1}}) &  &  & \mathrm{diag} (P^{N_G -1}_{\alpha})
  \end{array}\right)
\end{eqnarray}
Z$_{2}$ index of $\tilde{H}_{\mathbf k}$ is calculated using algorithm for noninteracting systems \cite{Fukui:2007p38592}. According to \cite{Wang:2011p39810} it is identical to the interacting Z$_{2}$ index of the original model.

\begin{figure}[t]
\begin{center}
\includegraphics[width=0.5\textwidth]{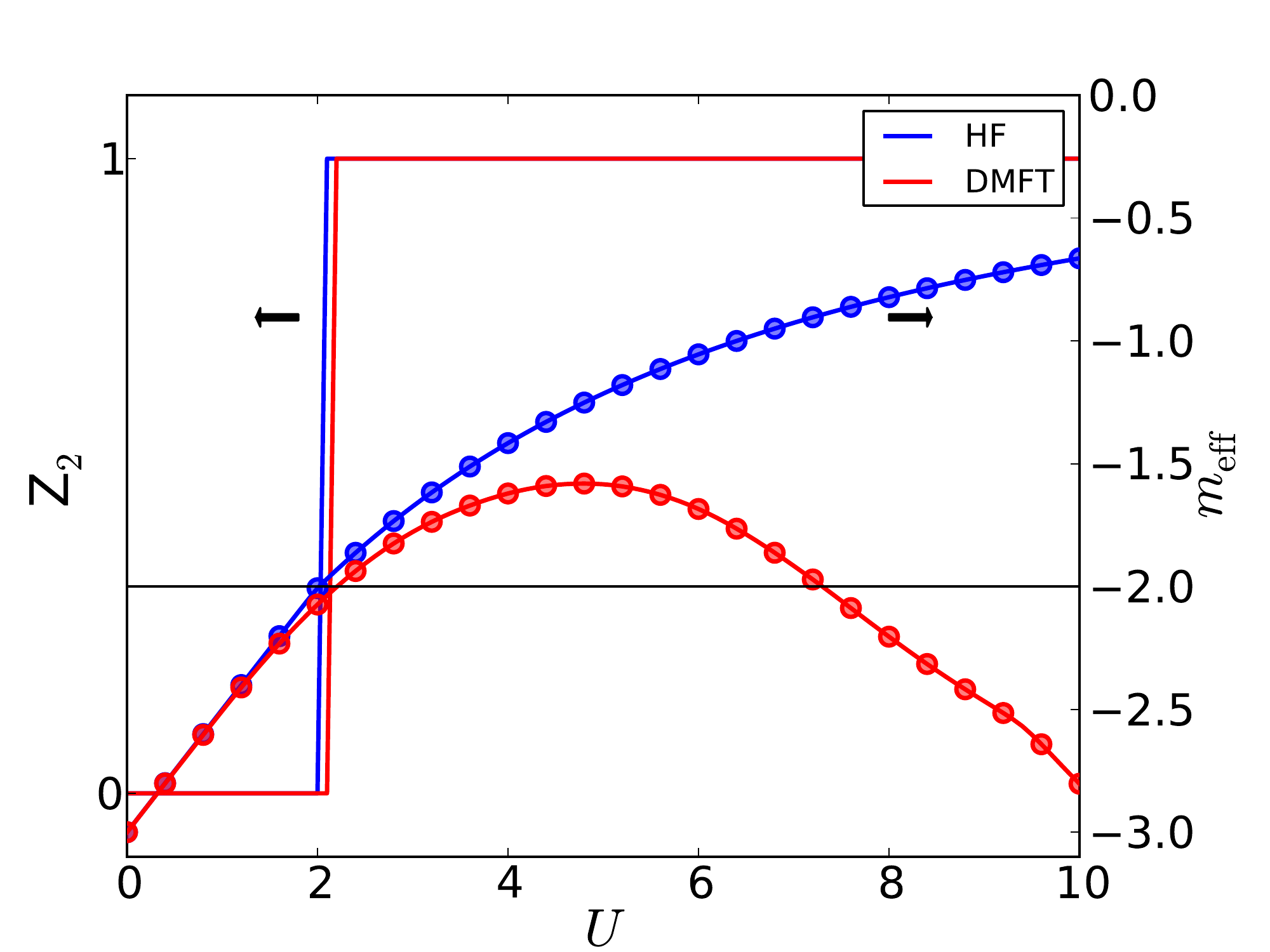}
\caption{Z$_{2}$ index (solid lines) and effective mass (line with dots) from Hartree-Fock and DMFT calculations. Z$_{2}$ index in HF treatment is inferred from value of $m_\mathrm{eff}$. In DMFT treatment, Z$_{2}$ index is calculated with information on poles of the local self-energies. It is also consistent with the edge spectra, see Fig.\ref{fig:edge}}
\label{fig:meff}
\end{center}
\end{figure}

\textit{Results--}In Fig.\ref{fig:meff} we plot the interacting Z$_{2}$ index (red line) as a function of interaction strength, it shows a jump from $0$ (trivial) to $1$ (nontrivial) at $U=2.2$, \textit{i.e.}, indicating an interaction induced topological phase transition.

We then validate and explore the nature of this topological transition by considering other physical quantities. First, we note that the static part of the self-energy $\Sigma_{\alpha}{(\infty)}= \frac{U(\langle n_{s}\rangle-\langle n_{p}\rangle)} {2}  \mathbb  {I}_{2}\otimes \sigma ^{z}$. It renormalizes parameter $m$ in $H_{0}$ to $m_\mathrm{eff} = m+U(\langle n_{s}\rangle-\langle n_{p}\rangle)/2$. In Fig.\ref{fig:meff} we plot the effective mass as a function of $U$. It can be seen that the topological phase transition occurs whenever $m_\mathrm{eff}$ exceeds $-2$. Topological transition occurs due to interaction induced relative level shift of $s$ and $p$ orbitals, which drives the system into an effective inverted band region.

Effective mass shows non-monotonous behavior for larger $U$. This is due to the fact that in the large $U$ limit fillings of $s$ and $p$ orbital tend to be evened out to $(1^{+},1^{-}) $. Since $\langle n_{s}\rangle-\langle n_{p}\rangle$ decays faster then $1/U$, $m_\mathrm{eff}$ turns back to the non-interacting value $m$ (see following). Interestingly, the interacting Z$_{2}$ index calculated in this region disassociates with effective mass. The system is always shows nontrivial topological property even as $m_\mathrm{eff}$ goes back to its noninteracting value. However, simple consideration based solely on effective mass would predict a trivial phase in this region. Above analysis shows that dynamical part of self-energy is important in determining the topological properties in a correlated topological insulator.


\begin{figure}[t]
\begin{center}
\includegraphics[width=0.4\textwidth]{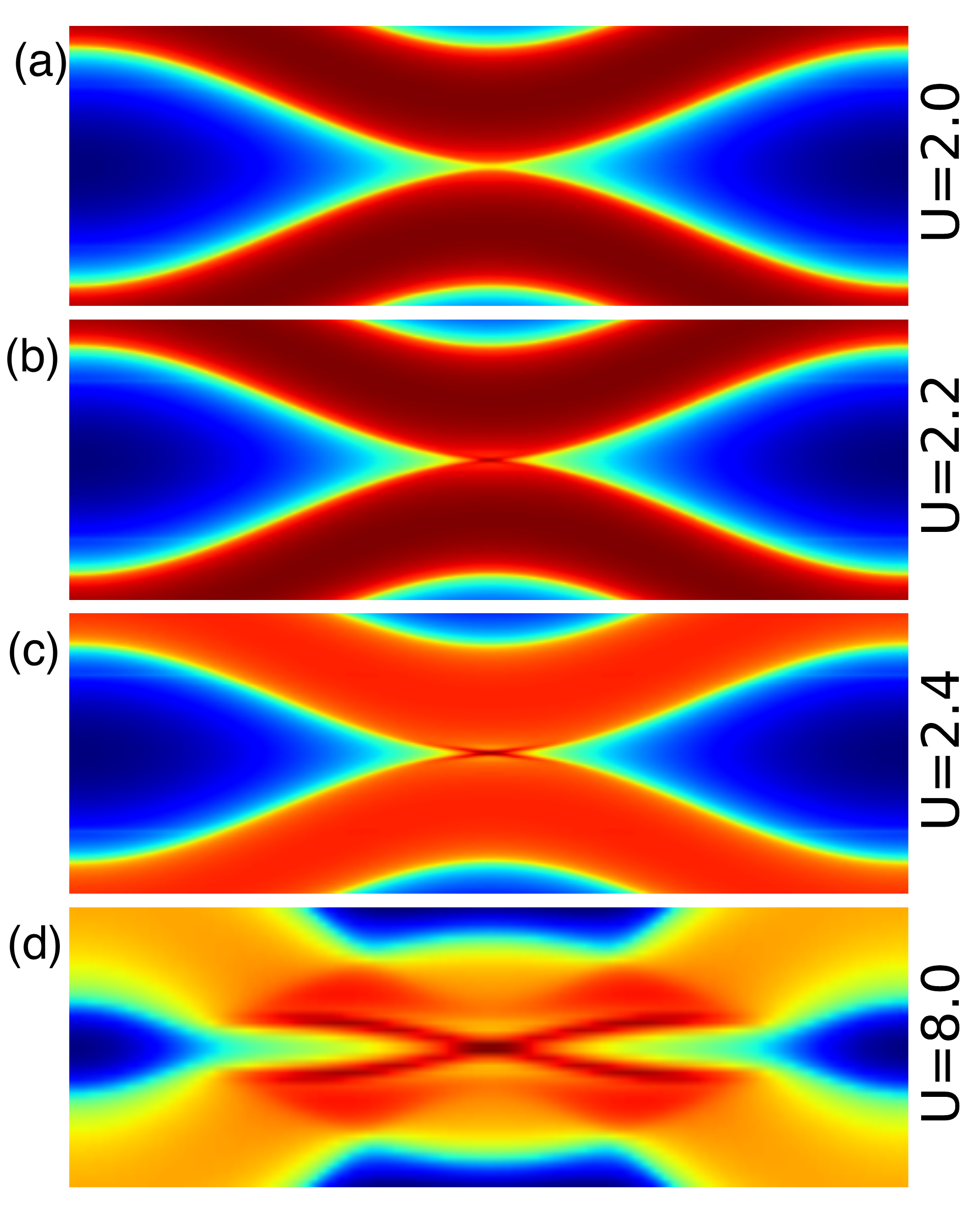}
\caption{Edge spectra for different interacting strengths. Across the topological transition point the edge state develops. Edge spectrum for $U=8.0$ still shows edge state crossing the fermi surface, indicating a nontrivial correlated TI phase. We show different frequency regions for $U=2.0,2.2,2.4$, $\omega\in[-2.4, 2.4]$ and for $U=8.0$, $\omega\in[-0.48, 0.48]$, respectively. }
\label{fig:edge}
\end{center}
\end{figure}

We further calculate edge spectrum $A(k_{x},\omega)$ by projecting the interaction Green's function on to the open edges \footnote{When calculating edge spectrum, we assume self-energy on the edge are the same with bulk, this is of course a crude approximation. It would be interesting to study position dependent (although local) self-energies in the formalism of real-space DMFT\cite{Potthoff:1999p4947,Potthoff:1999p4948}. }. In Fig.\ref{fig:edge}(a-c), edge states are in accordance with Z$_{2}$ index calculated from pole-expansions (TI phase for $U>2.2$). Inside TI phase, bulk gap first increases (due to band inversion) then shrinks (due to correlation effect) with increasing $U$.  Since the bulk gap never closes after the first transition at $U=2.2$, it is expected that the topological index is unchanged. For $U=8.0$, edge state is still visible, but the bulk gap is very small\ref{fig:edge}(d). Physically, this tiny gapped TI phase corresponds to a state proximate to Mott transition with a tiny quasi-particle weight. Due to the presence of crystal field splitting, the system does not reach to the case with $\langle n_{s}\rangle=\langle n_{p}\rangle=1$, thus preventing the Mott transition to happen. As long as there is a finite quasi-particle weight, the system is topological nontrivial. Reduction of the bulk gap for a correlated insulator is also observed in \cite{Yu:2011p35101,Sentef:2009p17333}.

As a comparison, we solve interacting BHZ model with the Hartree-Fock (HF) mean-field theory. In HF treatment, interacting term is decoupled as $H_{1} = U\sum_{i\sigma}( n_{s\sigma}\langle n_{s\bar{\sigma}}\rangle + n_{p\sigma} \langle n_{p\bar{\sigma}}\rangle) $. This also effectively modifies the mass term to $m_\mathrm{eff} = m + U(\langle n_{s}\rangle-\langle n_{p}\rangle)/2$. By solving the mean-field Hamiltonian self-consistently we find that $m_\mathrm{eff}$ increases monotonously with $U$. In Fig.\ref{fig:meff}, we see effective mass from DMFT and HF calculations are in accordance with each other for small interacting strength $U<2$. Interaction driven band inversion is captured by a Hatree-Fock level shift. However HF result deviates from DMFT one at large $U$. This is because that the HF theory does not correctly capture the asymptotic behavior of $\langle n_{s}\rangle-\langle n_{p}\rangle$ at large $U$ \footnote{A simple estimation based on two-level model shows that $\langle n_{s}\rangle-\langle n_{p}\rangle\sim1/U^{3}$, however, HF predicts $\langle n_{s}\rangle-\langle n_{p}\rangle\sim1/U$. }. Due to this, we anticipate that HF will favor the existence of topological phase transitions in some strongly correlated materials.

In HF treatment, topological properties is simply determined by the mean-field Hamiltonian, \textit{i.e.},  the mean-field Z$_{2}$ index comes from $m_\mathrm{eff}$. The topological transition occurs whenever $m_\mathrm{eff}$ exceeds $-2$. Topological transition point in HF theory shifts towards (slightly) smaller value of $U$ compared to the DMFT result. At the phase boundary, the single-particle excitation gap closes and the effective mass shows a kink in the transition point\cite{Cai:2008p20033, Wang:2010p25724}.

To summarize our physical result: the mean-field effect first drives the band inversion and turns the system into a correlated TI. Further increasing $U$ will first stabilize the TI phase by a larger band inversion, but soon correlation effect takes over to suppress the quasi-particle weight and hence the bulk gap. The system is academically always in the TI phase with a tiny gap since there is no Mott transition.


%
%
%
%
%
%


\textit{Discussion--}We have applied several ways to determine the topological properties of an interacting model, namely getting the interacting Z$_{2}$ index from pole-expansion, looking at the edge spectrum and using the mean-field theory. The mean-field theory is not reliable for strongly correlated systems. Practically, the edge spectrum (real frequency) is not directly accessible to many numerical techniques due to the complication of analytical continuation. And it is hard to distinguish odd and even pairs of edge states from the spectrum. The interacting Z$_{2}$ index Eqn.\ref{eqn:ishikawa} is defined using the Matsubara Green's function. However, as discussed in \cite{Wang:2011p43509, Wang:2011p39810} due to extremely inhomogeneity distribution of the integrand, it is hard to directly accomplish the integration to get quantized topological index. Pole expansion technique, combined with DMFT calculations, thus provides a practical way of determining the topological properties of correlated topological insulators.
Since pole-expansion technique is already widely used in LDA+DMFT calculations (for avoiding time consuming numerical inverse)\cite{Savrasov:2006p3872, Zhao:2011p43243}, with the proof-of-principle demonstration here, we anticipate that the combination of present approach with  the \textit{ab initio} tools will be useful for future theoretical searching of correlated TI materials.


\textit{Summary--} For parameter region studied in the present paper, the topological phase transition is essentially driven by mean-field effect (shift of HF levels). However, deep inside the interacting TI phase, correlation effect plays an important role in modifying the properties of the system. Mean-field theory fails in prediction of topological nature base on the orbital occupations. On the other hand, the interacting Z$_{2}$ index from the pole-expansion correctly predicts a topological nature of the correlated TI phase for whole range of interacting strength. This is quite remarkable, because we do not need to consult to dynamical quantities (like the edge spectrum or the real-frequency Green's function) to determine the topological nature of an interacting system.

\textit{Note--} After completion of this manuscript, there appear interesting papers by Wang \cite{Wang:2012p48720, Wang:2012p50546}.  We have validated their formula with our numerical Green's functions \footnote{We diagonalize $G({\mathbf k}_{i},0) =[-H_{\mathbf k}-\Sigma({\mathbf k}_{i},0)]^{-1}$ for ${\mathbf k}_{i}\in$ TRIM points. Denote eigenvectors with positive eigenvalues as $|\alpha_{i}\rangle$. They are also eigenvectors of parity operator $P=diag(1,1,-1,-1)$, \textit{i.e.} $P|\alpha_{i}\rangle=\eta_{\alpha_{i}}|\alpha_{i}\rangle$.
Then $[\frac{1}{2\pi i}\sum_{i,\alpha_{i}}
\ln\eta_{\alpha_{i}}]\mod 2$ gives Z$_{2}$ index. ($0$ for trivial insulator, $1$ for topological insulator.). }. It gives identical topological transition point as the pole-expansion method used in this paper. Independent studies are also reported in \cite{WitczakKrempa:2012p49698} and \cite{Budich:2012p51247}.

\textit{Acknowledgment--} This work is supported by NSF-China and National Basic Research Program of China (No.2009CB92102). We thank X.-L Qi for helpful discussions. LW thanks Matthias Troyer for generous support.
\bibliography{/Users/wanglei/Documents/Papers/papers}

\begin{thebibliography}{30}
\expandafter\ifx\csname natexlab\endcsname\relax\def\natexlab#1{#1}\fi
\expandafter\ifx\csname bibnamefont\endcsname\relax
  \def\bibnamefont#1{#1}\fi
\expandafter\ifx\csname bibfnamefont\endcsname\relax
  \def\bibfnamefont#1{#1}\fi
\expandafter\ifx\csname citenamefont\endcsname\relax
  \def\citenamefont#1{#1}\fi
\expandafter\ifx\csname url\endcsname\relax
  \def\url#1{\texttt{#1}}\fi
\expandafter\ifx\csname urlprefix\endcsname\relax\def\urlprefix{URL }\fi
\providecommand{\bibinfo}[2]{#2}
\providecommand{\eprint}[2][]{\url{#2}}

\bibitem[{\citenamefont{Wang et~al.}(2011{\natexlab{a}})\citenamefont{Wang,
  Jiang, Dai, and Xie}}]{Wang:2011p39810}
\bibinfo{author}{\bibfnamefont{L.}~\bibnamefont{Wang}},
  \bibinfo{author}{\bibfnamefont{H.}~\bibnamefont{Jiang}},
  \bibinfo{author}{\bibfnamefont{X.}~\bibnamefont{Dai}}, \bibnamefont{and}
  \bibinfo{author}{\bibfnamefont{X.~C.} \bibnamefont{Xie}},
  \bibinfo{journal}{arXiv} \textbf{\bibinfo{volume}{cond-mat.str-el}}
  (\bibinfo{year}{2011}{\natexlab{a}}), \eprint{1109.6292v1},
  \urlprefix\url{http://arxiv.org/abs/1109.6292v1}.

\bibitem[{\citenamefont{Zhang et~al.}(2009)\citenamefont{Zhang, Liu, Qi, Dai,
  Fang, and Zhang}}]{Zhang:2009p7135}
\bibinfo{author}{\bibfnamefont{H.-J.} \bibnamefont{Zhang}},
  \bibinfo{author}{\bibfnamefont{C.-X.} \bibnamefont{Liu}},
  \bibinfo{author}{\bibfnamefont{X.-L.} \bibnamefont{Qi}},
  \bibinfo{author}{\bibfnamefont{X.}~\bibnamefont{Dai}},
  \bibinfo{author}{\bibfnamefont{Z.}~\bibnamefont{Fang}}, \bibnamefont{and}
  \bibinfo{author}{\bibfnamefont{S.-C.} \bibnamefont{Zhang}},
  \bibinfo{journal}{Nat Phys} \textbf{\bibinfo{volume}{5}},
  \bibinfo{pages}{438} (\bibinfo{year}{2009}),
  \urlprefix\url{http://dx.doi.org/10.1038/nphys1270}.

\bibitem[{\citenamefont{Zhang et~al.}(2011{\natexlab{a}})\citenamefont{Zhang,
  Yu, Feng, Yao, Weng, Dai, and Fang}}]{Zhang:2011p30949}
\bibinfo{author}{\bibfnamefont{W.}~\bibnamefont{Zhang}},
  \bibinfo{author}{\bibfnamefont{R.}~\bibnamefont{Yu}},
  \bibinfo{author}{\bibfnamefont{W.}~\bibnamefont{Feng}},
  \bibinfo{author}{\bibfnamefont{Y.}~\bibnamefont{Yao}},
  \bibinfo{author}{\bibfnamefont{H.}~\bibnamefont{Weng}},
  \bibinfo{author}{\bibfnamefont{X.}~\bibnamefont{Dai}}, \bibnamefont{and}
  \bibinfo{author}{\bibfnamefont{Z.}~\bibnamefont{Fang}},
  \bibinfo{journal}{Phys. Rev. Lett.} \textbf{\bibinfo{volume}{106}},
  \bibinfo{pages}{156808} (\bibinfo{year}{2011}{\natexlab{a}}).

\bibitem[{\citenamefont{Xiao et~al.}(2010)\citenamefont{Xiao, Yao, Feng, Wen,
  Zhu, Chen, Stocks, and Zhang}}]{Xiao:2010p36216}
\bibinfo{author}{\bibfnamefont{D.}~\bibnamefont{Xiao}},
  \bibinfo{author}{\bibfnamefont{Y.}~\bibnamefont{Yao}},
  \bibinfo{author}{\bibfnamefont{W.}~\bibnamefont{Feng}},
  \bibinfo{author}{\bibfnamefont{J.}~\bibnamefont{Wen}},
  \bibinfo{author}{\bibfnamefont{W.}~\bibnamefont{Zhu}},
  \bibinfo{author}{\bibfnamefont{X.-Q.} \bibnamefont{Chen}},
  \bibinfo{author}{\bibfnamefont{G.}~\bibnamefont{Stocks}}, \bibnamefont{and}
  \bibinfo{author}{\bibfnamefont{Z.}~\bibnamefont{Zhang}},
  \bibinfo{journal}{Phys. Rev. Lett.} \textbf{\bibinfo{volume}{105}},
  \bibinfo{pages}{096404} (\bibinfo{year}{2010}).

\bibitem[{\citenamefont{Zhang et~al.}(2011{\natexlab{b}})\citenamefont{Zhang,
  Zhang, Weng, Zhang, Yang, Liu, Feng, Wang, Yu, Cao
  et~al.}}]{Zhang:2011p28394}
\bibinfo{author}{\bibfnamefont{J.~L.} \bibnamefont{Zhang}},
  \bibinfo{author}{\bibfnamefont{S.~J.} \bibnamefont{Zhang}},
  \bibinfo{author}{\bibfnamefont{H.}~\bibnamefont{Weng}},
  \bibinfo{author}{\bibfnamefont{W.}~\bibnamefont{Zhang}},
  \bibinfo{author}{\bibfnamefont{L.~X.} \bibnamefont{Yang}},
  \bibinfo{author}{\bibfnamefont{Q.~Q.} \bibnamefont{Liu}},
  \bibinfo{author}{\bibfnamefont{S.~M.} \bibnamefont{Feng}},
  \bibinfo{author}{\bibfnamefont{X.~C.} \bibnamefont{Wang}},
  \bibinfo{author}{\bibfnamefont{R.~C.} \bibnamefont{Yu}},
  \bibinfo{author}{\bibfnamefont{L.~Z.} \bibnamefont{Cao}},
  \bibnamefont{et~al.}, \bibinfo{journal}{Proceedings of the National Academy
  of Sciences of the United States of America} \textbf{\bibinfo{volume}{108}},
  \bibinfo{pages}{24} (\bibinfo{year}{2011}{\natexlab{b}}).

\bibitem[{\citenamefont{Varney et~al.}(2010)\citenamefont{Varney, Sun, Rigol,
  and Galitski}}]{Varney:2010p21916}
\bibinfo{author}{\bibfnamefont{C.~N.} \bibnamefont{Varney}},
  \bibinfo{author}{\bibfnamefont{K.}~\bibnamefont{Sun}},
  \bibinfo{author}{\bibfnamefont{M.}~\bibnamefont{Rigol}}, \bibnamefont{and}
  \bibinfo{author}{\bibfnamefont{V.}~\bibnamefont{Galitski}},
  \bibinfo{journal}{Phys Rev B} \textbf{\bibinfo{volume}{82}},
  \bibinfo{pages}{115125} (\bibinfo{year}{2010}).

\bibitem[{\citenamefont{Yu et~al.}(2011)\citenamefont{Yu, Xie, and
  Li}}]{Yu:2011p35101}
\bibinfo{author}{\bibfnamefont{S.-L.} \bibnamefont{Yu}},
  \bibinfo{author}{\bibfnamefont{X.-C.} \bibnamefont{Xie}}, \bibnamefont{and}
  \bibinfo{author}{\bibfnamefont{J.-X.} \bibnamefont{Li}},
  \bibinfo{journal}{Phys. Rev. Lett.} \textbf{\bibinfo{volume}{107}},
  \bibinfo{pages}{010401} (\bibinfo{year}{2011}).

\bibitem[{\citenamefont{Hohenadler et~al.}(2011)\citenamefont{Hohenadler, Lang,
  and Assaad}}]{Hohenadler:2011p28789}
\bibinfo{author}{\bibfnamefont{M.}~\bibnamefont{Hohenadler}},
  \bibinfo{author}{\bibfnamefont{T.~C.} \bibnamefont{Lang}}, \bibnamefont{and}
  \bibinfo{author}{\bibfnamefont{F.~F.} \bibnamefont{Assaad}},
  \bibinfo{journal}{Phys. Rev. Lett.} \textbf{\bibinfo{volume}{106}},
  \bibinfo{pages}{100403} (\bibinfo{year}{2011}).

\bibitem[{\citenamefont{Yoshida et~al.}(2011)\citenamefont{Yoshida, Fujimoto,
  and Kawakami}}]{Yoshida:2011p44360}
\bibinfo{author}{\bibfnamefont{T.}~\bibnamefont{Yoshida}},
  \bibinfo{author}{\bibfnamefont{S.}~\bibnamefont{Fujimoto}}, \bibnamefont{and}
  \bibinfo{author}{\bibfnamefont{N.}~\bibnamefont{Kawakami}},
  \bibinfo{journal}{arXiv} \textbf{\bibinfo{volume}{cond-mat.str-el}}
  (\bibinfo{year}{2011}), \eprint{1111.6250v1},
  \urlprefix\url{http://arxiv.org/abs/1111.6250v1}.

\bibitem[{\citenamefont{Wang et~al.}(2010{\natexlab{a}})\citenamefont{Wang,
  Shi, Zhang, Wang, Dai, and Xie}}]{Wang:2010p25724}
\bibinfo{author}{\bibfnamefont{L.}~\bibnamefont{Wang}},
  \bibinfo{author}{\bibfnamefont{H.}~\bibnamefont{Shi}},
  \bibinfo{author}{\bibfnamefont{S.}~\bibnamefont{Zhang}},
  \bibinfo{author}{\bibfnamefont{X.}~\bibnamefont{Wang}},
  \bibinfo{author}{\bibfnamefont{X.}~\bibnamefont{Dai}}, \bibnamefont{and}
  \bibinfo{author}{\bibfnamefont{X.-C.} \bibnamefont{Xie}},
  \bibinfo{journal}{arXiv} \textbf{\bibinfo{volume}{cond-mat.str-el}}
  (\bibinfo{year}{2010}{\natexlab{a}}), \eprint{1012.5163v1},
  \urlprefix\url{http://arxiv.org/abs/1012.5163v1}.

\bibitem[{\citenamefont{Rachel and LeHur}(2010)}]{Rachel:2010p20458}
\bibinfo{author}{\bibfnamefont{S.}~\bibnamefont{Rachel}} \bibnamefont{and}
  \bibinfo{author}{\bibfnamefont{K.}~\bibnamefont{LeHur}},
  \bibinfo{journal}{Phys Rev B} \textbf{\bibinfo{volume}{82}},
  \bibinfo{pages}{075106} (\bibinfo{year}{2010}).

\bibitem[{\citenamefont{Varney et~al.}(2011)\citenamefont{Varney, Sun, Rigol,
  and Galitski}}]{Varney:2011p45724}
\bibinfo{author}{\bibfnamefont{C.}~\bibnamefont{Varney}},
  \bibinfo{author}{\bibfnamefont{K.}~\bibnamefont{Sun}},
  \bibinfo{author}{\bibfnamefont{M.}~\bibnamefont{Rigol}}, \bibnamefont{and}
  \bibinfo{author}{\bibfnamefont{V.}~\bibnamefont{Galitski}},
  \bibinfo{journal}{Physical Review B} \textbf{\bibinfo{volume}{84}},
  \bibinfo{pages}{241105} (\bibinfo{year}{2011}).

\bibitem[{\citenamefont{Niu et~al.}(1985)\citenamefont{Niu, Thouless, and
  Wu}}]{Niu:1985p19667}
\bibinfo{author}{\bibfnamefont{Q.}~\bibnamefont{Niu}},
  \bibinfo{author}{\bibfnamefont{D.~J.} \bibnamefont{Thouless}},
  \bibnamefont{and} \bibinfo{author}{\bibfnamefont{Y.-S.} \bibnamefont{Wu}},
  \bibinfo{journal}{Phys Rev B} \textbf{\bibinfo{volume}{31}},
  \bibinfo{pages}{3372} (\bibinfo{year}{1985}).

\bibitem[{\citenamefont{Wang et~al.}(2010{\natexlab{b}})\citenamefont{Wang, Qi,
  and Zhang}}]{Wang:2010p25171}
\bibinfo{author}{\bibfnamefont{Z.}~\bibnamefont{Wang}},
  \bibinfo{author}{\bibfnamefont{X.-L.} \bibnamefont{Qi}}, \bibnamefont{and}
  \bibinfo{author}{\bibfnamefont{S.-C.} \bibnamefont{Zhang}},
  \bibinfo{journal}{Phys. Rev. Lett.} \textbf{\bibinfo{volume}{105}},
  \bibinfo{pages}{256803} (\bibinfo{year}{2010}{\natexlab{b}}).

\bibitem[{\citenamefont{Wang et~al.}(2011{\natexlab{b}})\citenamefont{Wang,
  Dai, and Xie}}]{Wang:2011p43509}
\bibinfo{author}{\bibfnamefont{L.}~\bibnamefont{Wang}},
  \bibinfo{author}{\bibfnamefont{X.}~\bibnamefont{Dai}}, \bibnamefont{and}
  \bibinfo{author}{\bibfnamefont{X.~C.} \bibnamefont{Xie}},
  \bibinfo{journal}{Physical Review B} \textbf{\bibinfo{volume}{84}},
  \bibinfo{pages}{205116} (\bibinfo{year}{2011}{\natexlab{b}}).

\bibitem[{\citenamefont{Bernevig et~al.}(2006)\citenamefont{Bernevig, Hughes,
  and Zhang}}]{Bernevig:2006p44981}
\bibinfo{author}{\bibfnamefont{B.~A.} \bibnamefont{Bernevig}},
  \bibinfo{author}{\bibfnamefont{T.~L.} \bibnamefont{Hughes}},
  \bibnamefont{and} \bibinfo{author}{\bibfnamefont{S.-C.} \bibnamefont{Zhang}},
  \bibinfo{journal}{Science} \textbf{\bibinfo{volume}{314}},
  \bibinfo{pages}{1757} (\bibinfo{year}{2006}).

\bibitem[{\citenamefont{Qi et~al.}(2008)\citenamefont{Qi, Hughes, and
  Zhang}}]{Qi:2008p12545}
\bibinfo{author}{\bibfnamefont{X.-L.} \bibnamefont{Qi}},
  \bibinfo{author}{\bibfnamefont{T.~L.} \bibnamefont{Hughes}},
  \bibnamefont{and} \bibinfo{author}{\bibfnamefont{S.-C.} \bibnamefont{Zhang}},
  \bibinfo{journal}{Phys Rev B} \textbf{\bibinfo{volume}{78}},
  \bibinfo{pages}{195424} (\bibinfo{year}{2008}).

\bibitem[{\citenamefont{Georges et~al.}(1996)\citenamefont{Georges, Kotliar,
  Krauth, and Rozenberg}}]{Georges:1996p5571}
\bibinfo{author}{\bibfnamefont{A.}~\bibnamefont{Georges}},
  \bibinfo{author}{\bibfnamefont{G.}~\bibnamefont{Kotliar}},
  \bibinfo{author}{\bibfnamefont{W.}~\bibnamefont{Krauth}}, \bibnamefont{and}
  \bibinfo{author}{\bibfnamefont{M.~J.} \bibnamefont{Rozenberg}},
  \bibinfo{journal}{Rev Mod Phys} \textbf{\bibinfo{volume}{68}},
  \bibinfo{pages}{13} (\bibinfo{year}{1996}).

\bibitem[{\citenamefont{Balzer et~al.}(2011)\citenamefont{Balzer, Gdaniec, and
  Potthoff}}]{Balzer:2011p44597}
\bibinfo{author}{\bibfnamefont{M.}~\bibnamefont{Balzer}},
  \bibinfo{author}{\bibfnamefont{N.}~\bibnamefont{Gdaniec}}, \bibnamefont{and}
  \bibinfo{author}{\bibfnamefont{M.}~\bibnamefont{Potthoff}},
  \bibinfo{journal}{arXiv} \textbf{\bibinfo{volume}{cond-mat.str-el}}
  (\bibinfo{year}{2011}), \eprint{1109.1205v2},
  \urlprefix\url{http://arxiv.org/abs/1109.1205v2}.

\bibitem[{\citenamefont{Zhao et~al.}(2011)\citenamefont{Zhao, Zhuang, Deng,
  Cai, Fang, and Dai}}]{Zhao:2011p43243}
\bibinfo{author}{\bibfnamefont{J.}~\bibnamefont{Zhao}},
  \bibinfo{author}{\bibfnamefont{J.-N.} \bibnamefont{Zhuang}},
  \bibinfo{author}{\bibfnamefont{X.}~\bibnamefont{Deng}},
  \bibinfo{author}{\bibfnamefont{L.}~\bibnamefont{Cai}},
  \bibinfo{author}{\bibfnamefont{Z.}~\bibnamefont{Fang}}, \bibnamefont{and}
  \bibinfo{author}{\bibfnamefont{X.}~\bibnamefont{Dai}},
  \bibinfo{journal}{arXiv} \textbf{\bibinfo{volume}{cond-mat.str-el}}
  (\bibinfo{year}{2011}), \eprint{1111.2157v1},
  \urlprefix\url{http://arxiv.org/abs/1111.2157v1}.

\bibitem[{\citenamefont{Savrasov et~al.}(2006)\citenamefont{Savrasov, Haule,
  and Kotliar}}]{Savrasov:2006p3872}
\bibinfo{author}{\bibfnamefont{S.}~\bibnamefont{Savrasov}},
  \bibinfo{author}{\bibfnamefont{K.}~\bibnamefont{Haule}}, \bibnamefont{and}
  \bibinfo{author}{\bibfnamefont{G.}~\bibnamefont{Kotliar}},
  \bibinfo{journal}{Phys. Rev. Lett.} \textbf{\bibinfo{volume}{96}},
  \bibinfo{pages}{036404} (\bibinfo{year}{2006}).

\bibitem[{\citenamefont{Fukui and Hatsugai}(2007)}]{Fukui:2007p38592}
\bibinfo{author}{\bibfnamefont{T.}~\bibnamefont{Fukui}} \bibnamefont{and}
  \bibinfo{author}{\bibfnamefont{Y.}~\bibnamefont{Hatsugai}},
  \bibinfo{journal}{J. Phys. Soc. Jpn.} \textbf{\bibinfo{volume}{76}},
  \bibinfo{pages}{053702} (\bibinfo{year}{2007}).

\bibitem[{\citenamefont{Sentef et~al.}(2009)\citenamefont{Sentef, Kune{\v s},
  Werner, and Kampf}}]{Sentef:2009p17333}
\bibinfo{author}{\bibfnamefont{M.}~\bibnamefont{Sentef}},
  \bibinfo{author}{\bibfnamefont{J.}~\bibnamefont{Kune{\v s}}},
  \bibinfo{author}{\bibfnamefont{P.}~\bibnamefont{Werner}}, \bibnamefont{and}
  \bibinfo{author}{\bibfnamefont{A.}~\bibnamefont{Kampf}},
  \bibinfo{journal}{Phys Rev B} \textbf{\bibinfo{volume}{80}},
  \bibinfo{pages}{155116} (\bibinfo{year}{2009}).

\bibitem[{\citenamefont{Cai et~al.}(2008)\citenamefont{Cai, Chen, Kou, and
  Wang}}]{Cai:2008p20033}
\bibinfo{author}{\bibfnamefont{Z.}~\bibnamefont{Cai}},
  \bibinfo{author}{\bibfnamefont{S.}~\bibnamefont{Chen}},
  \bibinfo{author}{\bibfnamefont{S.}~\bibnamefont{Kou}}, \bibnamefont{and}
  \bibinfo{author}{\bibfnamefont{Y.}~\bibnamefont{Wang}},
  \bibinfo{journal}{Phys Rev B} \textbf{\bibinfo{volume}{78}},
  \bibinfo{pages}{035123} (\bibinfo{year}{2008}).

\bibitem[{\citenamefont{Wang et~al.}(2012)\citenamefont{Wang, Qi, and
  Zhang}}]{Wang:2012p48720}
\bibinfo{author}{\bibfnamefont{Z.}~\bibnamefont{Wang}},
  \bibinfo{author}{\bibfnamefont{X.-L.} \bibnamefont{Qi}}, \bibnamefont{and}
  \bibinfo{author}{\bibfnamefont{S.-C.} \bibnamefont{Zhang}},
  \bibinfo{journal}{arXiv} \textbf{\bibinfo{volume}{cond-mat.str-el}}
  (\bibinfo{year}{2012}), \eprint{1201.6431v2},
  \urlprefix\url{http://arxiv.org/abs/1201.6431v2}.

\bibitem[{\citenamefont{Wang and Zhang}(2012)}]{Wang:2012p50546}
\bibinfo{author}{\bibfnamefont{Z.}~\bibnamefont{Wang}} \bibnamefont{and}
  \bibinfo{author}{\bibfnamefont{S.-C.} \bibnamefont{Zhang}},
  \bibinfo{journal}{arXiv} \textbf{\bibinfo{volume}{cond-mat.str-el}}
  (\bibinfo{year}{2012}), \eprint{1203.1028v2},
  \urlprefix\url{http://arxiv.org/abs/1203.1028v2}.

\bibitem[{\citenamefont{Witczak-Krempa
  et~al.}(2012)\citenamefont{Witczak-Krempa, Jeon, Park, and
  Kim}}]{WitczakKrempa:2012p49698}
\bibinfo{author}{\bibfnamefont{W.}~\bibnamefont{Witczak-Krempa}},
  \bibinfo{author}{\bibfnamefont{G.~S.} \bibnamefont{Jeon}},
  \bibinfo{author}{\bibfnamefont{K.}~\bibnamefont{Park}}, \bibnamefont{and}
  \bibinfo{author}{\bibfnamefont{Y.~B.} \bibnamefont{Kim}},
  \bibinfo{journal}{arXiv} \textbf{\bibinfo{volume}{cond-mat.str-el}}
  (\bibinfo{year}{2012}), \eprint{1202.4460v1},
  \urlprefix\url{http://arxiv.org/abs/1202.4460v1}.

\bibitem[{\citenamefont{Budich et~al.}(2012)\citenamefont{Budich, Thomale,
  Laubach, and Zhang}}]{Budich:2012p51247}
\bibinfo{author}{\bibfnamefont{J.~C.} \bibnamefont{Budich}},
  \bibinfo{author}{\bibfnamefont{R.}~\bibnamefont{Thomale}},
  \bibinfo{author}{\bibfnamefont{M.}~\bibnamefont{Laubach}}, \bibnamefont{and}
  \bibinfo{author}{\bibfnamefont{S.-C.} \bibnamefont{Zhang}},
  \bibinfo{journal}{arXiv} \textbf{\bibinfo{volume}{cond-mat.str-el}}
  (\bibinfo{year}{2012}), \eprint{1203.2928v1},
  \urlprefix\url{http://arxiv.org/abs/1203.2928v1}.

\bibitem[{\citenamefont{Potthoff and
  Nolting}(1999{\natexlab{a}})}]{Potthoff:1999p4947}
\bibinfo{author}{\bibfnamefont{M.}~\bibnamefont{Potthoff}} \bibnamefont{and}
  \bibinfo{author}{\bibfnamefont{W.}~\bibnamefont{Nolting}},
  \bibinfo{journal}{Phys Rev B} \textbf{\bibinfo{volume}{60}},
  \bibinfo{pages}{7834} (\bibinfo{year}{1999}{\natexlab{a}}).

\bibitem[{\citenamefont{Potthoff and
  Nolting}(1999{\natexlab{b}})}]{Potthoff:1999p4948}
\bibinfo{author}{\bibfnamefont{M.}~\bibnamefont{Potthoff}} \bibnamefont{and}
  \bibinfo{author}{\bibfnamefont{W.}~\bibnamefont{Nolting}},
  \bibinfo{journal}{Phys Rev B} \textbf{\bibinfo{volume}{59}},
  \bibinfo{pages}{2549} (\bibinfo{year}{1999}{\natexlab{b}}).

\end{thebibliography}
\end{document}